\documentclass[journal,12pt]{IEEEtran}
\ifCLASSOPTIONcompsoc
  \usepackage[nocompress]{cite}
\else
  \usepackage{cite}
\fi
\ifCLASSINFOpdf
\else
\fi

\usepackage{hyperref}
\usepackage{url}
\usepackage{graphicx}
\usepackage{booktabs} 
\usepackage{adjustbox}
\usepackage{amsmath}
\usepackage{colortbl} 
\usepackage{xcolor} 
\usepackage[skins]{tcolorbox}
\usepackage[ruled,vlined,noresetcount]{algorithm2e}
\usepackage{courier}

\begin{document}

\title{Text-Independent Speaker Verification Using Long Short-Term Memory Networks}

\author{\IEEEauthorblockN{Aryan Mobiny}\\
\IEEEauthorblockA{Department of Electrical and Computer Engineering,\\
University of Houston,\\
Houston, Texas 77004,\\
Email: amobiny@uh.edu
}

}

\author{
    \IEEEauthorblockN{Aryan Mobiny\IEEEauthorrefmark{1}, Mohammad Najarian\IEEEauthorrefmark{2}}\\
    \IEEEauthorblockA{\IEEEauthorrefmark{1} \small Department of Electrical and Computer Engineering, University of Houston\\
\texttt{\small Email: amobiny@uh.edu}}\\
    \IEEEauthorblockA{\IEEEauthorrefmark{2}Department of Industrial Engineering, University of Houston\\
\texttt{Email: mnajarian@uh.edu}
    }
}

\maketitle

\begin{abstract}

In this paper, an architecture based on Long Short-Term Memory Networks has been proposed for the text-independent scenario which is aimed to capture the temporal speaker-related information by operating over traditional speech features. For speaker verification, at first, a background model must be created for speaker representation. Then, in enrollment stage, the speaker models will be created based on the enrollment utterances. For this work, the model will be trained in an end-to-end fashion to combine the first two stages. The main goal of end-to-end training is the model being optimized to be consistent with the speaker verification protocol. The end-to-end training jointly learns the background and speaker models by creating the representation space. The LSTM architecture is trained to create a discrimination space for validating the match and non-match pairs for speaker verification. The proposed architecture demonstrate its superiority in the text-independent compared to other traditional methods.
 
\end{abstract}

\IEEEpeerreviewmaketitle

\section{Introduction}

The main goal of Speaker Verification~(SV) is the process of verifying a query sample belonging to a speaker utterance by comparing to the existing speaker models. Speaker verification is usually split into two text-independent and text-dependant categories. Text-dependent includes the scenario in which all the speakers are uttering the same phrase while in text-independent no prior information is considered for what the speakers are saying. The later setting is much more challenging as it can contain numerous variations for non-speaker information that can be misleading while extracting solely speaker information is desired.

The speaker verification, in general, consists of three stages: Training, enrollment, and evaluation. In training, the universal background model is trained using the gallery of speakers. In enrollment, based on the created background model, the new speakers will be enrolled in creating the speaker model. Technically, the speakers' models are generated using the universal background model. In the evaluation phase, the test utterances will be compared to the speaker models for further identification or verification.


Recently, by the success of deep learning in applications such as in biomedical purposes ~\cite{shen2015multi, mobiny2017lung}, automatic speech recognition, image recognition and network sparsity~\cite{simonyan2014very,krizhevsky2012imagenet,hinton2012deep,torfi2018attention}, the DNN-based approaches have also been proposed for Speaker Recognition~(SR)~\cite{lei2014novel,variani2014deep}. 
%

The traditional speaker verification models such as Gaussian Mixture Model-Universal Background Model (GMM-UBM) \cite{reynolds2000speaker} and i-vector \cite{dehak2011front} have been the state-of-the-art for long. The drawback of these approaches is the employed unsupervised fashion that does not optimize them for verification setup. Recently, supervised methods proposed for model adaptation to speaker verification such as the one presented in \cite{campbell2006support} and PLDA-based i-vectors model~\cite{garcia2011analysis}. Convolutional Neural Networks~(CNNs) has also been used for speech recognition and speaker verification~\cite{variani2014deep,abdel2014convolutional} inspired by their their superior power for action recognition~\cite{ji20133d} and scene understanding~\cite{tran2015learning}. Capsule networks introduced by Hinton et al.~\cite{sabour2017dynamic} has shown quite remarkable performance in different tasks ~\cite{mobiny2018fast, jaiswal2018capsulegan}, and demonstrated the potential and power to be used for similar purposes.

In the present work, we propose the use of LSTMs by using MFCCs\footnote{Mel Frequency Cepstral Coefficients} speech features for directly capturing the temporal information of the speaker-related information rather than dealing with non-speaker information which plays no role for speaker verification.

\section{Related works}

There is a huge literature on speaker verification. However, we only focus on the research efforts which are based on deep learning deep learning. One of the traditional successful works in speaker verification is the use of Locally Connected Networks (LCNs)~\cite{chen2015locally} for the text-dependent scenario. Deep networks have also been used as feature extractor for representing speaker models~\cite{heigold2016end,zhang2017end}. We investigate LSTMs in an end-to-end fashion for speaker verification. As Convolutional Neural Networks \cite{lecun1998gradient} have successfully been used for the speech recognition~\cite{sainath2013deep} some works use their architecture for speaker verification~\cite{lei2014novel,richardson2015deep}. The most similar work to ours is \cite{heigold2016end} in which they use LSTMs for the text-dependent setting. On the contrary, we use LSTMs for the text-independent scenario which is a more challenging one.

\section{Speaker Verification Using Deep Neural Networks}

Here, we explain the speaker verification phases using deep learning. In different works, these steps have been adopted regarding the procedure proposed by their research efforts such as i-vector~\cite{dehak2011front,kenny2007joint}, d-vector system~\cite{variani2014deep}.\\

\subsection{Development} 

In the development stage which also called training, the speaker utterances are used for background model generation which ideally should be a universal model for speaker model representation. DNNs are employed due to their power for feature extraction. By using deep models, the feature learning will be done for creating an output space which represents the speaker in a universal model.\\

\subsection{Enrollment}

In this phase, a model must be created for each speaker. For each speaker, by collecting the spoken utterances and feeding to the trained network, different output features will be generated for speaker utterances. From this point, different approaches have been proposed on how to integrate these enrollment features for creating the speaker model. The tradition one is aggregating the representations by averaging the outputs of the DNN which is called d-vector system~\cite{chen2015locally,variani2014deep}.
\subsection{Evaluation}

For evaluation, the test utterance is the input of the network and the output is the utterance representative. The output representative will be compared to different speaker model and the verification criterion will be some similarity function. For evaluation purposes, the traditional Equal Error Rate~(EER) will often be used which is the operating point in that false reject rate and false accept rate are equal.

\section{Model}

The main goal is to implement LSTMs on top of speech extracted features. The input to the model as well as the architecture itself is explained in the following subsections.

\subsection{Input}

\begin{figure}[ht]
\begin{center}
\includegraphics[scale=0.45]{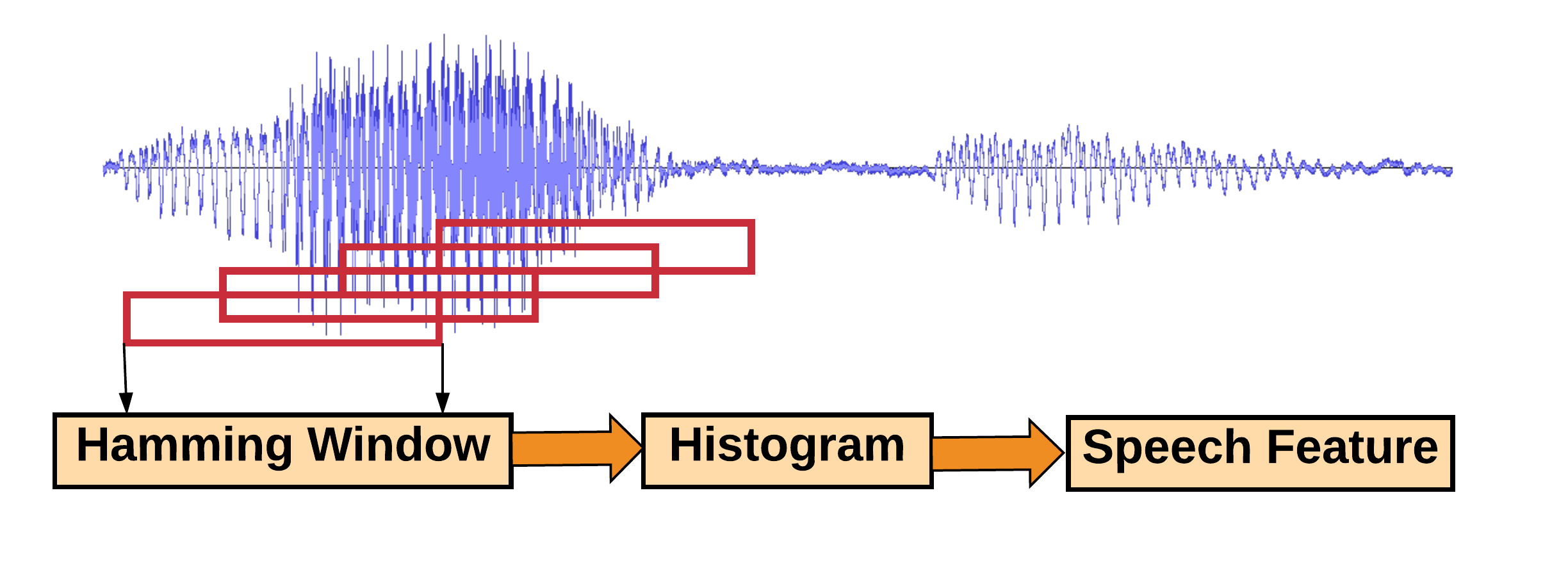}
\end{center}
\caption{The feature extraction from the raw signal.}
\label{fig:speechinput}
\end{figure}

The raw signal is extracted and 25ms windows with \%60 overlapping are used for the generation of the spectrogram as depicted in Fig.~\ref{fig:speechinput}. By selecting 1-second of the sound stream, 40 log-energy of filter banks per window and performing mean and variance normalization, a feature window of $40 \times 100$ is generated for each 1-second utterance. Before feature extraction, voice activity detection has been done over the raw input for eliminating the silence. The derivative feature has not been used as using them did not make any improvement considering the empirical evaluations. For feature extraction, we used SpeechPy library~\cite{torfi2018speechpy}.

\subsection{Architecture}\label{sec:architecture}

\begin{figure}[ht]
\begin{center}
\includegraphics[scale=0.7]{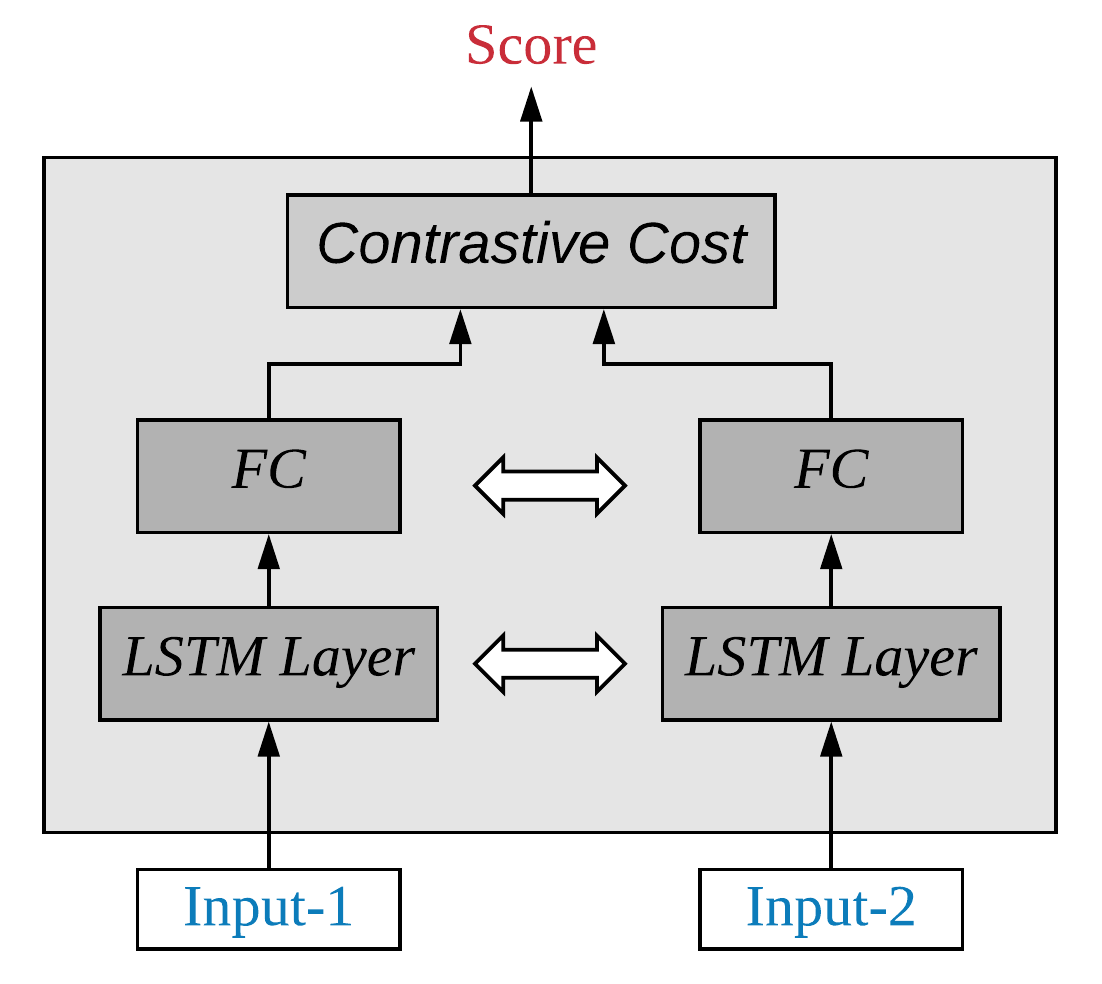}
\end{center}
\caption{The siamese architecture built based on two LSTM layers with weight sharing.}
\label{fig:lstm}
\end{figure}

The architecture that we use a long short-term memory recurrent neural network (LSTM)~\cite{hochreiter1997long,sak2014long} with a single output for decision making. We input fixed-length sequences although LSTMs are not limited by this constraint. Only the last hidden state of the LSTM model is used for decision making using the loss function. The LSTM that we use has two layers with 300 nodes each~(Fig.~\ref{fig:lstm}).

\subsection{Verification Setup}\label{sec:Verification Setup}

A usual method which has been used in many other works~\cite{chen2015locally}, is training the network using the Softmax loss function for the auxiliary classification task and then use the extracted features for the main verification purpose. A reasonable argument about this approach is that the Softmax criterion is not in align with the verification protocol due to optimizing for identification of individuals and not the one-vs-one comparison. Technically, the Softmax optimization criterion is as below:

\begin{equation}
\textrm{softmax}({\bf x})_{Speaker} = \frac{e^{x_{Speaker}}}{\sum_{Dev_{Spk}}{e^{x_{Dev_{Spk}}}}}
\end{equation}

\begin{equation}\label{eq4}
  \begin{cases}
    x_{Speaker} = W_{Speaker} \times y + b\\
    x_{Dev_{Spk}} = W_{Dev_{Spk}} \times y + b
  \end{cases}
\end{equation}

\noindent in which \textit{Speaker} and \textit{$Dev_{Spk}$} denote the sample speaker and an identity from speaker development set, respectively. As it is clear from the criterion, there is no indication to the one-to-one speaker comparison for being consistent to speaker verification mode.

To consider this condition, we use the Siamese architecture to satisfy the verification purpose which has been proposed in \cite{chopra2005learning} and employed in different applications~\cite{sun2018deep,varior2016gated,koch2015siamese}. As we mentioned before, the Softmax optimization will be used for initialization and the obtained weights will be used for fine-tuning.

The Siamese architecture consists of two identical networks with weight sharing. The goal is to create a shared feature subspace which is aimed at discrimination between genuine and impostor pairs. The main idea is that when two elements of an input pair are from the same identity, their output distances should be close and far away, otherwise. For this objective, the training loss will be contrastive cost function. The aim of contrastive loss $C_W(X,Y)$ is the minimization of the loss in both scenarios of having genuine and impostor pairs, with the following definition:
\begin{align}\label{eq2}
C_W(X,Y) = {{1}\over{N}}  \sum_{j=1}^{N} C_W(Y_j,(X_{1},X_{2})_j),
\end{align}

\noindent where N indicates the training samples, $j$ is the sample index and $C_W(Y_i,(X_{p_{1}},X_{p_{2}})_i)$ will be defined as follows:

\begin{equation} \label{eq3}
\begin{split}
C_W&(Y_i,(X_{1},X_{2})_j) = Y*C_{gen}(D_W(X_{1},X_{2})_j)\\ &+ (1-Y)*C_{imp}(D_W(X_{1},X_{2})_j)+\lambda{||W||}_{2}
\end{split}
\end{equation}

\noindent in which the last term is the regularization. $C_{gen}$ and $C_{imp}$ will be defined as the functions of $D_W(X_{1},X_{2})$ by the following equations:

\begin{equation}\label{eq5}
  \begin{cases}
    C_{gen}(D_W(X_{1},X_{2})={{1}\over{2}}{D_W(X_{1},X_{2})}^2\\
    C_{imp}(D_W(X_{1},X_{2})={{1}\over{2}}max\{0,{(M-D_W(X_{1},X_{2}))}\}^2
  \end{cases}
\end{equation}

\noindent in which $M$ is the margin.

\section{Experiments}

TensorFLow has been used as the deep learning library~\cite{tensorflow2015-whitepaper}. For the development phase, we used data augmentation by randomly sampling the 1-second audio sample for each person at a time. Batch normalization has also been used for avoiding possible gradient explotion~\cite{ioffe2015batch}. It's been shown that effective pair selection can drastically improve the verification accuracy~\cite{lin2008facetnet}. Speaker verification is performed using the protocol consistent with \cite{Nagrani17} for which the name identities start with ’E’ will be used for evaluation. 

\begin{algorithm}\label{algorithm:pair selection} 
 \textbf{Update}: Freeze weights!
 
 \textbf{Evaluate}: Input data and get output distance vector\;
 \textbf{Search}: Return max and min distances for match pairs : $max\_gen$ \& $min\_gen$\;
 \textbf{Thresholding}: Calculate $th=th_{0}\times \left | \frac{max\_gen}{min\_gen}\right|$\;
 \While{impostor pair}{
  
  \eIf{$imp > max\_gen + th$}{
   discard\;
   }{
   feed the pair\;
  }
 }
 \caption{The utilized pair selection algorithm for selecting the main contributing impostor pairs}
\end{algorithm}

\subsection{Baselines}

We compare our method with different baseline methods. The \textit{GMM-UBM} method \cite{reynolds2000speaker} if the first candidate. The MFCCs features with 40 coefficients are extracted and used. The \textit{Universal Background Model}~(UBM) is trained using 1024 mixture components. The \textit{I-Vector} model~\cite{dehak2011front}, with and without \textit{Probabilistic Linear Discriminant Analysis}~(PLDA)~\cite{kenny2010bayesian}, has also been implemented as the baseline. 

The other baseline is the use of DNNs with locally-connected layers as proposed in \cite{chen2015locally}. In the d-vector system, after development phase, the d-vectors extracted from the enrollment utterances will be aggregated to each other for generating the final representation. Finally, in the evaluation stage, the similarity function determines the closest d-vector of the test utterances to the speaker models.

\subsection{Comparison to Different Methods}

Here we compare the baseline approaches with the proposed model as provided in Table ~\ref{table:compasison}. We utilized the architecture and the setup as discussed in Section~\ref{sec:architecture} and Section~\ref{sec:Verification Setup}, respectively. As can be seen in Table ~\ref{table:compasison}, our proposed architecture outperforms the other methods.  

\begin{table}[h]
\caption[Table caption text]{The architecture used for verification purpose.}
\label{table:compasison}
\begin{center}
\addtolength{\tabcolsep}{0pt}
\begin{tabular}{cc}
\toprule 
Model & EER\\
\hline
\midrule
\rowcolor{black!0} GMM-UBM~\cite{reynolds2000speaker} & 27.1 \\ 
\rowcolor{black!0} I-vectors~\cite{dehak2011front}   & 24.7 \\
\rowcolor{black!0} I-vectors~\cite{dehak2011front} + PLDA~\cite{kenny2010bayesian}  & 23.5 \\
\rowcolor{black!0} LSTM~[ours]  & 22.9 \\

\bottomrule
\end{tabular}
\end{center}

\end{table}

\subsection{Effect of Utterance Duration}

\begin{figure}[ht]
\begin{center}
\includegraphics[width=0.5\textwidth]{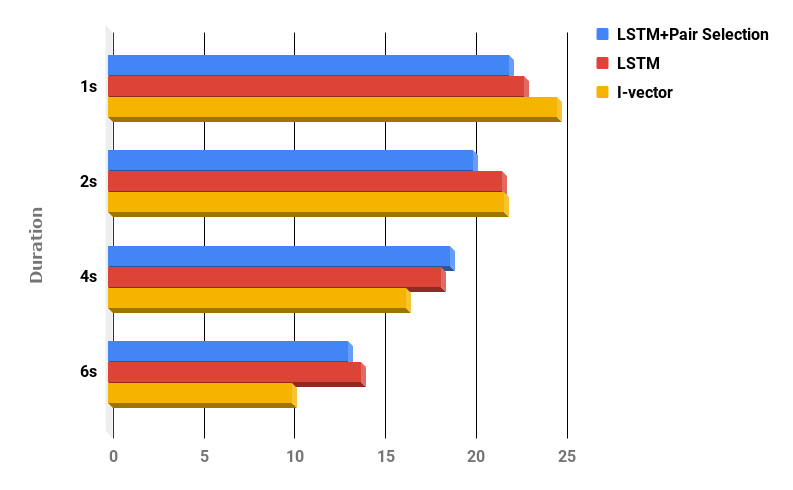}
\end{center}
\caption{The effect of the utterance duration~(EER).}
\label{fig:utteranceeffect}
\end{figure}

One one the main advantage of the baseline methods such as \cite{dehak2011front} is their ability to capture robust speaker characteristics through long utterances. As demonstrated in Fig.~\ref{fig:utteranceeffect}, our proposed method outperforms the others for short utterances considering we used 1-second utterances. However, it is worth to have a fair comparison for longer utterances as well. In order to have a one-to-one comparison, we modified our architecture to feed and train the system on longer utterances. In all experiments, the duration of utterances utilized for development, enrollment, and evaluation are the same.

%

As can be observed in Fig.~\ref{fig:utteranceeffect}, the superiority of our method is only in short utterances and in longer utterances, the traditional baseline methods such as \cite{dehak2011front}, still are the winners and LSTMs fails to capture effectively inter- and inter-speaker variations.

\section{Conclusion}

In this work, an end-to-end model based on LSTMs has been proposed for text-independent speaker verification. It was shown that the model provided promising results for capturing the temporal information in addition to capture the within-speaker information. The proposed LSTM architecture has directly been used on the speech features extracted from speaker utterances for modeling the spatiotemporal information. One the observed traces is the superiority of traditional methods on longer utterances for more robust speaker modeling. More rigorous studies are needed to investigate the reasoning behind the failure of LSTMs to capture long dependencies for speaker related characteristics. Additionally, it is expected that the combination of traditional models with long short-term memory architectures may improve the accuracy by capturing the long-term dependencies in a more effective way. The main advantage of the proposed approach is its ability to capture informative features in short utterances.

%
%


\bibliographystyle{ieeetr}
\bibliography{ref}

\begin{thebibliography}{10}

\bibitem{shen2015multi}
W.~Shen, M.~Zhou, F.~Yang, C.~Yang, and J.~Tian, ``Multi-scale convolutional
  neural networks for lung nodule classification,'' in {\em International
  Conference on Information Processing in Medical Imaging}, pp.~588--599,
  Springer, 2015.

\bibitem{mobiny2017lung}
A.~Mobiny, S.~Moulik, I.~Gurcan, T.~Shah, and H.~Van~Nguyen, ``Lung cancer
  screening using adaptive memory-augmented recurrent networks,'' {\em arXiv
  preprint arXiv:1710.05719}, 2017.

\bibitem{simonyan2014very}
K.~Simonyan and A.~Zisserman, ``Very deep convolutional networks for
  large-scale image recognition,'' {\em arXiv preprint arXiv:1409.1556}, 2014.

\bibitem{krizhevsky2012imagenet}
A.~Krizhevsky, I.~Sutskever, and G.~E. Hinton, ``Imagenet classification with
  deep convolutional neural networks,'' in {\em Advances in neural information
  processing systems}, pp.~1097--1105, 2012.

\bibitem{hinton2012deep}
G.~Hinton, L.~Deng, D.~Yu, G.~E. Dahl, A.-r. Mohamed, N.~Jaitly, A.~Senior,
  V.~Vanhoucke, P.~Nguyen, T.~N. Sainath, {\em et~al.}, ``Deep neural networks
  for acoustic modeling in speech recognition: The shared views of four
  research groups,'' {\em IEEE Signal Processing Magazine}, vol.~29, no.~6,
  pp.~82--97, 2012.

\bibitem{torfi2018attention}
A.~Torfi and R.~A. Shirvani, ``Attention-based guided structured sparsity of
  deep neural networks,'' {\em arXiv preprint arXiv:1802.09902}, 2018.

\bibitem{lei2014novel}
Y.~Lei, N.~Scheffer, L.~Ferrer, and M.~McLaren, ``A novel scheme for speaker
  recognition using a phonetically-aware deep neural network,'' in {\em
  Acoustics, Speech and Signal Processing (ICASSP), 2014 IEEE International
  Conference on}, pp.~1695--1699, IEEE, 2014.

\bibitem{variani2014deep}
E.~Variani, X.~Lei, E.~McDermott, I.~L. Moreno, and J.~Gonzalez-Dominguez,
  ``Deep neural networks for small footprint text-dependent speaker
  verification,'' in {\em Acoustics, Speech and Signal Processing (ICASSP),
  2014 IEEE International Conference on}, pp.~4052--4056, IEEE, 2014.

\bibitem{reynolds2000speaker}
D.~A. Reynolds, T.~F. Quatieri, and R.~B. Dunn, ``Speaker verification using
  adapted gaussian mixture models,'' {\em Digital signal processing}, vol.~10,
  no.~1-3, pp.~19--41, 2000.

\bibitem{dehak2011front}
N.~Dehak, P.~J. Kenny, R.~Dehak, P.~Dumouchel, and P.~Ouellet, ``Front-end
  factor analysis for speaker verification,'' {\em IEEE Transactions on Audio,
  Speech, and Language Processing}, vol.~19, no.~4, pp.~788--798, 2011.

\bibitem{campbell2006support}
W.~M. Campbell, D.~E. Sturim, and D.~A. Reynolds, ``Support vector machines
  using gmm supervectors for speaker verification,'' {\em IEEE signal
  processing letters}, vol.~13, no.~5, pp.~308--311, 2006.

\bibitem{garcia2011analysis}
D.~Garcia-Romero and C.~Y. Espy-Wilson, ``Analysis of i-vector length
  normalization in speaker recognition systems,'' in {\em Twelfth Annual
  Conference of the International Speech Communication Association}, 2011.

\bibitem{abdel2014convolutional}
O.~Abdel-Hamid, A.-r. Mohamed, H.~Jiang, L.~Deng, G.~Penn, and D.~Yu,
  ``Convolutional neural networks for speech recognition,'' {\em IEEE/ACM
  Transactions on audio, speech, and language processing}, vol.~22, no.~10,
  pp.~1533--1545, 2014.

\bibitem{ji20133d}
S.~Ji, W.~Xu, M.~Yang, and K.~Yu, ``3d convolutional neural networks for human
  action recognition,'' {\em IEEE transactions on pattern analysis and machine
  intelligence}, vol.~35, no.~1, pp.~221--231, 2013.

\bibitem{tran2015learning}
D.~Tran, L.~Bourdev, R.~Fergus, L.~Torresani, and M.~Paluri, ``Learning
  spatiotemporal features with 3d convolutional networks,'' in {\em Computer
  Vision (ICCV), 2015 IEEE International Conference on}, pp.~4489--4497, IEEE,
  2015.

\bibitem{sabour2017dynamic}
S.~Sabour, N.~Frosst, and G.~E. Hinton, ``Dynamic routing between capsules,''
  in {\em Advances in Neural Information Processing Systems}, pp.~3856--3866,
  2017.

\bibitem{mobiny2018fast}
A.~Mobiny and H.~Van~Nguyen, ``Fast capsnet for lung cancer screening,'' {\em
  arXiv preprint arXiv:1806.07416}, 2018.

\bibitem{jaiswal2018capsulegan}
A.~Jaiswal, W.~AbdAlmageed, and P.~Natarajan, ``Capsulegan: Generative
  adversarial capsule network,'' {\em arXiv preprint arXiv:1802.06167}, 2018.

\bibitem{chen2015locally}
Y.-h. Chen, I.~Lopez-Moreno, T.~N. Sainath, M.~Visontai, R.~Alvarez, and
  C.~Parada, ``Locally-connected and convolutional neural networks for small
  footprint speaker recognition,'' in {\em Sixteenth Annual Conference of the
  International Speech Communication Association}, 2015.

\bibitem{heigold2016end}
G.~Heigold, I.~Moreno, S.~Bengio, and N.~Shazeer, ``End-to-end text-dependent
  speaker verification,'' in {\em Acoustics, Speech and Signal Processing
  (ICASSP), 2016 IEEE International Conference on}, pp.~5115--5119, IEEE, 2016.

\bibitem{zhang2017end}
C.~Zhang and K.~Koishida, ``End-to-end text-independent speaker verification
  with triplet loss on short utterances,'' in {\em Proc. of Interspeech}, 2017.

\bibitem{lecun1998gradient}
Y.~LeCun, L.~Bottou, Y.~Bengio, and P.~Haffner, ``Gradient-based learning
  applied to document recognition,'' {\em Proceedings of the IEEE}, vol.~86,
  no.~11, pp.~2278--2324, 1998.

\bibitem{sainath2013deep}
T.~N. Sainath, A.-r. Mohamed, B.~Kingsbury, and B.~Ramabhadran, ``Deep
  convolutional neural networks for lvcsr,'' in {\em Acoustics, speech and
  signal processing (ICASSP), 2013 IEEE international conference on},
  pp.~8614--8618, IEEE, 2013.

\bibitem{richardson2015deep}
F.~Richardson, D.~Reynolds, and N.~Dehak, ``Deep neural network approaches to
  speaker and language recognition,'' {\em IEEE Signal Processing Letters},
  vol.~22, no.~10, pp.~1671--1675, 2015.

\bibitem{kenny2007joint}
P.~Kenny, G.~Boulianne, P.~Ouellet, and P.~Dumouchel, ``Joint factor analysis
  versus eigenchannels in speaker recognition,'' {\em IEEE Transactions on
  Audio, Speech, and Language Processing}, vol.~15, no.~4, pp.~1435--1447,
  2007.

\bibitem{torfi2018speechpy}
A.~Torfi, ``Speechpy-a library for speech processing and recognition,'' {\em
  arXiv preprint arXiv:1803.01094}, 2018.

\bibitem{hochreiter1997long}
S.~Hochreiter and J.~Schmidhuber, ``Long short-term memory,'' {\em Neural
  computation}, vol.~9, no.~8, pp.~1735--1780, 1997.

\bibitem{sak2014long}
H.~Sak, A.~Senior, and F.~Beaufays, ``Long short-term memory recurrent neural
  network architectures for large scale acoustic modeling,'' in {\em Fifteenth
  annual conference of the international speech communication association},
  2014.

\bibitem{chopra2005learning}
S.~Chopra, R.~Hadsell, and Y.~LeCun, ``Learning a similarity metric
  discriminatively, with application to face verification,'' in {\em Computer
  Vision and Pattern Recognition, 2005. CVPR 2005. IEEE Computer Society
  Conference on}, vol.~1, pp.~539--546, IEEE, 2005.

\bibitem{sun2018deep}
X.~Sun, A.~Torfi, and N.~Nasrabadi, ``Deep siamese convolutional neural
  networks for identical twins and look-alike identification,'' {\em Deep
  Learning in Biometrics}, p.~65, 2018.

\bibitem{varior2016gated}
R.~R. Varior, M.~Haloi, and G.~Wang, ``Gated siamese convolutional neural
  network architecture for human re-identification,'' in {\em European
  Conference on Computer Vision}, pp.~791--808, Springer, 2016.

\bibitem{koch2015siamese}
G.~Koch, R.~Zemel, and R.~Salakhutdinov, ``Siamese neural networks for one-shot
  image recognition,'' in {\em ICML Deep Learning Workshop}, vol.~2, 2015.

\bibitem{tensorflow2015-whitepaper}
M.~Abadi, A.~Agarwal, P.~Barham, E.~Brevdo, Z.~Chen, C.~Citro, G.~S. Corrado,
  A.~Davis, J.~Dean, M.~Devin, S.~Ghemawat, I.~Goodfellow, A.~Harp, G.~Irving,
  M.~Isard, Y.~Jia, R.~Jozefowicz, L.~Kaiser, M.~Kudlur, J.~Levenberg,
  D.~Man\'{e}, R.~Monga, S.~Moore, D.~Murray, C.~Olah, M.~Schuster, J.~Shlens,
  B.~Steiner, I.~Sutskever, K.~Talwar, P.~Tucker, V.~Vanhoucke, V.~Vasudevan,
  F.~Vi\'{e}gas, O.~Vinyals, P.~Warden, M.~Wattenberg, M.~Wicke, Y.~Yu, and
  X.~Zheng, ``{TensorFlow}: Large-scale machine learning on heterogeneous
  systems,'' 2015.
\newblock Software available from tensorflow.org.

\bibitem{ioffe2015batch}
S.~Ioffe and C.~Szegedy, ``Batch normalization: Accelerating deep network
  training by reducing internal covariate shift,'' in {\em International
  conference on machine learning}, pp.~448--456, 2015.

\bibitem{lin2008facetnet}
Y.-R. Lin, Y.~Chi, S.~Zhu, H.~Sundaram, and B.~L. Tseng, ``Facetnet: a
  framework for analyzing communities and their evolutions in dynamic
  networks,'' in {\em Proceedings of the 17th international conference on World
  Wide Web}, pp.~685--694, ACM, 2008.

\bibitem{Nagrani17}
A.~Nagrani, J.~S. Chung, and A.~Zisserman, ``Voxceleb: a large-scale speaker
  identification dataset,'' in {\em INTERSPEECH}, 2017.

\bibitem{kenny2010bayesian}
P.~Kenny, ``Bayesian speaker verification with heavy-tailed priors.,'' in {\em
  Odyssey}, p.~14, 2010.

\end{thebibliography}
\end{document}